\documentclass[journal=chreay,layout=traditional,manuscript=review]{achemso}
\usepackage{color,lscape}
\usepackage{mathrsfs}
\usepackage{array}
\usepackage{caption}
\usepackage{verbatim}
\usepackage{amsfonts, amssymb,amsmath}
\usepackage{graphicx}
\usepackage{comment}

\title{Filopodia rotate and coil by actively generating twist in their actin shaft}

\author{Natascha Leijnse$^{**}$}
\affiliation{Niels Bohr Institute, University of Copenhagen, 2100 Copenhagen, Denmark}

\author{Younes Farhangi Barooji$^{**}$}
\affiliation{Niels Bohr Institute, University of Copenhagen, 2100 Copenhagen, Denmark}

\author{Mohammad Reza Arastoo}
\affiliation{Niels Bohr Institute, University of Copenhagen, 2100 Copenhagen, Denmark}

\author{Stine Lauritzen S\o nder}
 \affiliation{Membrane Integrity, Danish Cancer Society Research Center, Strandboulevarden 49, 2100 Copenhagen, Denmark}

\author{Bram Verhagen}
\affiliation{Niels Bohr Institute, University of Copenhagen, 2100 Copenhagen, Denmark}

\author{Lena Wullkopf}
 \affiliation{Biotech Research and Innovation Centre (BRIC),  University of Copenhagen, Ole Maal\o es Vej 5, 2200 Copenhagen, Denmark}

\author{Janine Terra Erler}
 \affiliation{Biotech Research and Innovation Centre (BRIC), University of Copenhagen, Ole Maal\o es Vej 5, 2200 Copenhagen, Denmark}
 
\author{Szabolcs Semsey}
\affiliation{Niels Bohr Institute, University of Copenhagen, 2100 Copenhagen, Denmark}

\author{Jesper Nylandsted}
 \affiliation{Membrane Integrity, Danish Cancer Society Research Center, Strandboulevarden 49, 2100 Copenhagen, Denmark and Department of Cellular and Molecular Medicine, Faculty of Health Sciences, University of Copenhagen, Blegdamsvej 3C, 2200 Copenhagen, Denmark}
 
\author{Lene Broeng Oddershede}
\affiliation{Niels Bohr Institute, University of Copenhagen, 2100 Copenhagen, Denmark}

\author{Amin Doostmohammadi}
\affiliation{Niels Bohr Institute, University of Copenhagen, 2100 Copenhagen, Denmark}
\email{theory: doostmohammadi@nbi.ku.dk}

\author{Poul Martin Bendix}
\affiliation{Niels Bohr Institute, University of Copenhagen, 2100 Copenhagen, Denmark}
\email{experiments: bendix@nbi.ku.dk}

\begin{document}

\maketitle
\clearpage

\section*{Abstract}
Filopodia are actin-rich structures, present on the surface of practically every known eukaryotic cell. These structures play a pivotal role in specific cell-cell and cell-matrix interactions by allowing cells to explore their environment, generate mechanical forces, perform chemical signaling, or convey signals via intercellular tunneling nano-bridges. The dynamics of filopodia appear quite complex as they exhibit a rich behavior of buckling, pulling, length and shape changes. Here, we show that filopodia additionally explore their 3D extracellular space by combining growth and shrinking with axial twisting and buckling of their actin rich core. Importantly, the actin core inside filopodia performs a twisting or spinning motion which is observed for a range of highly distinct and cognate cell types spanning from earliest development to highly differentiated tissue cells. Non-equilibrium physical modeling of actin and myosin confirm that twist, and hence rotation, is an emergent phenomenon of active filaments confined in a narrow channel which points to a generic mechanism present in all cells. Our measurements confirm that filopodia exert traction forces and form helical buckles in a range of different cell types that can be ascribed to accumulation of sufficient twist. These results lead us to conclude that activity induced twisting of the actin shaft is a general mechanism underlying fundamental functions of filopodia.\newline
\textbf{** Shared first authorship}
\textbf{* Corresponding author: Experiments: bendix@nbi.ku.dk; Theory: doostmohammadi@nbi.ku.dk}

\section*{Introduction}
Mechanical interactions between cells and their environment are essential for cellular functions like motility, communication, and sensing. The initial contact formed by cells is mediated by F-actin rich filopodia which are highly dynamic tubular structures on the cell surface that allow cells to reach out and interact with their extracellular environment and adjacent cells~\cite{mattila}.\par

Filopodia are present in a wide variety of cell types ranging from embryonic stem cells~\cite{sanders} to neuronal cells~\cite{bornschlogl1,leijnse1,lowery,tamada}, they are important for cell migration during wound healing~\cite{wood} and in cellular disorders such as cancer~\cite{arjonen}. Recently, filopodia have been discovered to play critical roles in development by facilitating communication between mesenchymal stem cells~\cite{sanders} and during compaction of the early embryo~\cite{fierrogonzalez}.

Structurally, filopodia are formed as thin membrane protrusions (diameter between $100$ and $300$~nm) containing $10$-$30$~bundles of actin filaments, which are cross-linked by molecules such as fascin~\cite{vignjevic}. Transmembrane integrins link the actin to the cell membrane while peripheral proteins like IBAR link the actin to the inner leaflet of the plasma membrane~\cite{mattila}.

Filopodia can differ significantly in length from a few micrometers to tens of micrometers. Long specialized filopodia can be sub-categorized into: Cytonemes~\cite{gonzalezmendez2019} which are involved in long range cell signalling; tunneling nanotubes~\cite{sanders} involved in intercellular material exchange including cell-cell virus transmission~\cite{lehmann}; and recently discovered airinemes~\cite{eom2020} found on skin resident macrophages involved in pigment pattern formation during zebrafish development.


Despite the high diversity of mechanical functions carried out by filopodia there seem to be common characteristics which are preserved in all types of filopodia. These include typical traction forces of tens of piconewtons~\cite{chan,moller,murphy, leijnse2, bornschlogl2}, growth and shrinkage~\cite{bornschlogl1}, and bending or lever arm activity~\cite{kress}. 

Growth and shrinkage of filopodia are regulated by actin polymerization at the tip~\cite{mallavarapu} and myosin activity which contributes to retrograde flow~\cite{bornschlogl2,aratyn}. In addition to this, a sweeping motion of the filopodial tip around the cellular base has been reported~\cite{bornschlogl1,zidovska} and even rotational motion has been indicated in \textcolor{black}{HEK293} cells~\cite{leijnse2} and neuronal cells~\cite{tamada}. However, these studies have not been able to decipher whether the filamentous actin inside the filopodium, in the following denoted as Ôactin shaftÕ, performs a spinning motion around its own axis or whether the filopodium performs a circular sweeping motion around the anchor point. Complex movements and helical buckling shapes have been reported together with simultaneous force generation~\cite{leijnse2,zidovska,tamada,nemethova}, thus indicating build-up of torsional twist in the actin shaft of \textcolor{black}{HEK293} cell filopodia. The dynamics of filopodial movement has been shown to be unaffected by inhibition of myosin II \cite{nemethova}. Knock-down of myosin Va and Vb, which are motors walking in a spiral path on actin filaments, was found to reduce the lateral movements of filopodia tips~\cite{ali2002,nishizaka,tamada}. However, the generality and  underlying mechanisms for all these modes of movement has remained enigmatic.\par

Here we show that twist generation in the actin shaft can explain several of the observed phenomena such as helical buckling, traction, and rotational movements of filopodia. We developed a novel optical tweezers/confocal microscopy assay to visualize the rotation of the F-actin shaft inside a filopodium. An optical trap is used to fix the filopodial tip and also used to attach a bead to the side of a filopodium. This bead linked to the actin inside the filopodium via vitronectin-integrin bonds and hence reports about any axial rotation and retrograde flow of actin. The observed axial rotations of the bead are compared to results from tracking of filopodial tips of cells grown on glass or embedded in a collagen I matrix. Furthermore, helical buckles resulting from twist accumulation are detected in a number of cell types, on glass as well as embedded in collagen I. To test whether these buckles originate from membrane induced compressional load~\cite{pronk} or from accumulation of twist within the actin shaft resulting from the spinning, we quantify the forces acting along the actin shaft. The generality of these phenomena was assessed by measuring the traction forces of both naive pluripotent stem cells and terminally differentiated cells. Interestingly, despite the presence of traction forces delivered by filopodia in all cell types, we frequently observed helical buckling in the actin shaft which can be explained by build-up of torsional twist.\par

We identify a mechanism behind the twisting and consequent spinning of the actin core by modeling actin/myosin complexes inside the filopodial membrane as an active gel, showing that mirror-symmetry breaking and the emergence of twist are generic phenomena in a confined assembly of active filaments, suggesting the physical origin of the observed twisting motion in variety of cell types.

\section*{Results}
\label{sec:headings}

\subsection*{Filopodia and extracted membrane tethers can rotate, pull, and buckle}

\begin{figure}[h]
 \centering
  \includegraphics[scale=0.5]{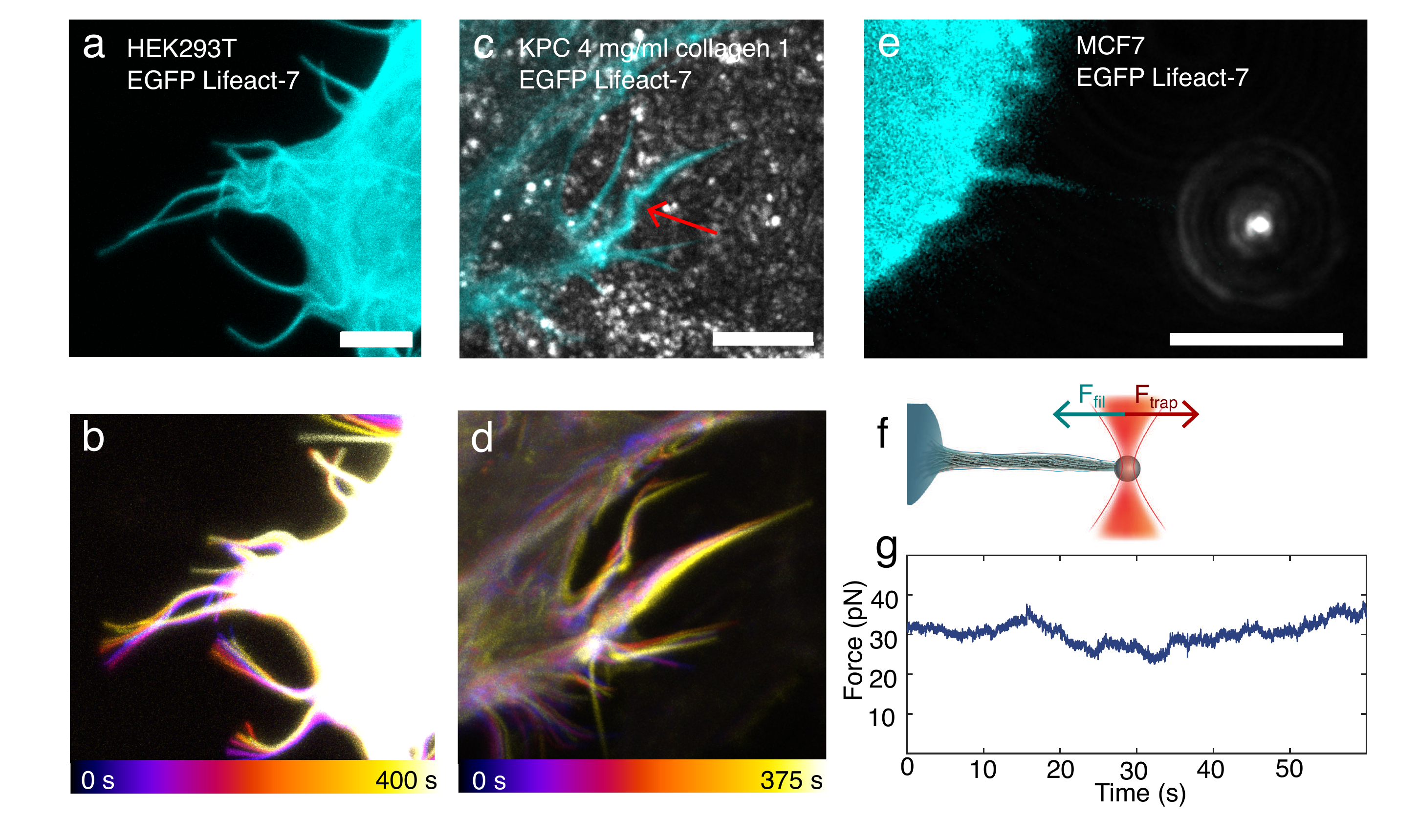}
\caption{{\bf Free and confined filopodia show rich dynamics that can lead to bending, buckling, coiling, shortening, and pulling.}
(a and b) The tips of free filopodia rotate. Confocal $Z$-stacks of a HEK293T cell (cyan, EGFP Lifeact-7) grown on glass. (a) Confocal $Z$-projection of a single $Z$-stack. Scale bar is $5$~$\mu$m. (b) Overlay of $Z$-projections at $11$ consecutive time points (total time is $400$~s, color coded from blue to white).
(c and d) Filopodia confined by collagen  frequently exhibit buckles. Confocal $Z$-stacks of a KPC cell (cyan, EGFP Lifeact-7) grown in $4$~mg/ml collagen I (grey, reflection). The red \textcolor{black}{arrow marks} filopodial bending regions.
(c) Confocal $Z$-projection of a single $Z$-stack. Scale bar is $5$~$\mu$m. (d) Overlay of $Z$-projections at $6$ consecutive time points (total time is $375$~s, color coded from blue to white.
(e, f, g) Membrane tethers extracted from cells fill up with F-actin (actin shaft), are highly dynamic, and behave like filopodia. (e) Confocal image of a tether extracted from a MCF7 cell (cyan, EGFP Lifeact-7) using an optically trapped vitronectin coated $d=4.95$~$\mu$m bead (grey, reflection). Scale bar is $5$~$\mu$m. (f) Schematics of the setup for tether extraction: The force exerted by the optical trap (orange laser beam profile) holds a bead (grey) in place. This bead is used to extract a membrane tether from the cell (cyan). The tether exerts a pulling force on the bead (green arrow) towards the cell body, counteracted by the trap force (red arrow). The force exerted by the tether is measured by tracking the beadÕs displacement relative to the initial center of the trap. Over time, the tether begins to bend, coil, and hence shortens and exerts a traction force on the trapped bead. (g) Holding force as a function of time exerted by the tether from (e) versus time.}
\label{fig:fig1}
\end{figure}

Filopodia show rich and complicated dynamics that is connected to remodeling of their actin shaft. On the order of minutes, they undergo significant movement and reshaping. Filopodial tips are often found to undergo a rotation~\cite{leijnse2,zidovska,tamada,leijnse3} or a sweeping  motion as shown for a EGFP Lifeact-7 expressing HEK293T cell on glass in Fig.~\ref{fig:fig1}a and b and Supplementary Movie 1a. Bending or coiling are frequently observed phenomena, shown in  Fig.~\ref{fig:fig1}c and d for a KP$^{\text{R172H}}$C cell (cyan, mEmerald Lifeact-7, \textcolor{black}{in the following denoted as KPC}) embedded in a matrix containing $4$~mg/ml collagen I (Supplementary Fig. 1 and 2 and Supplementary Movie 1b). The traction force generated by filopodia is examined by extracting a new filopodium from the plasma membrane using an optically trapped bead (Fig.~\ref{fig:fig1}e-g), and the trapping of the tip also reveals frequent formation of buckles observed along the filopodium (Supplementary Movie 1c). Buckle formation additionally took place without prior imaging with the confocal scanning laser, thus excluding that buckling is an artifact of the cell reacting to the laser illumination, see Supplementary Fig.~3.\par

The rotating motion of a filopodium and the associated shape changes and generation of traction, shown in Fig. \ref{fig:fig1}, indicate that the actin shaft contains twist that can be converted into buckling and traction. We therefore next sought to investigate possible generation of torque by an acto-myosin system confined within a tether extracted from a living cell. 

\subsection*{The actin shaft within extracted membrane tethers performs a spinning motion}

\begin{figure}[h]
 \centering
  \includegraphics[scale=0.5]{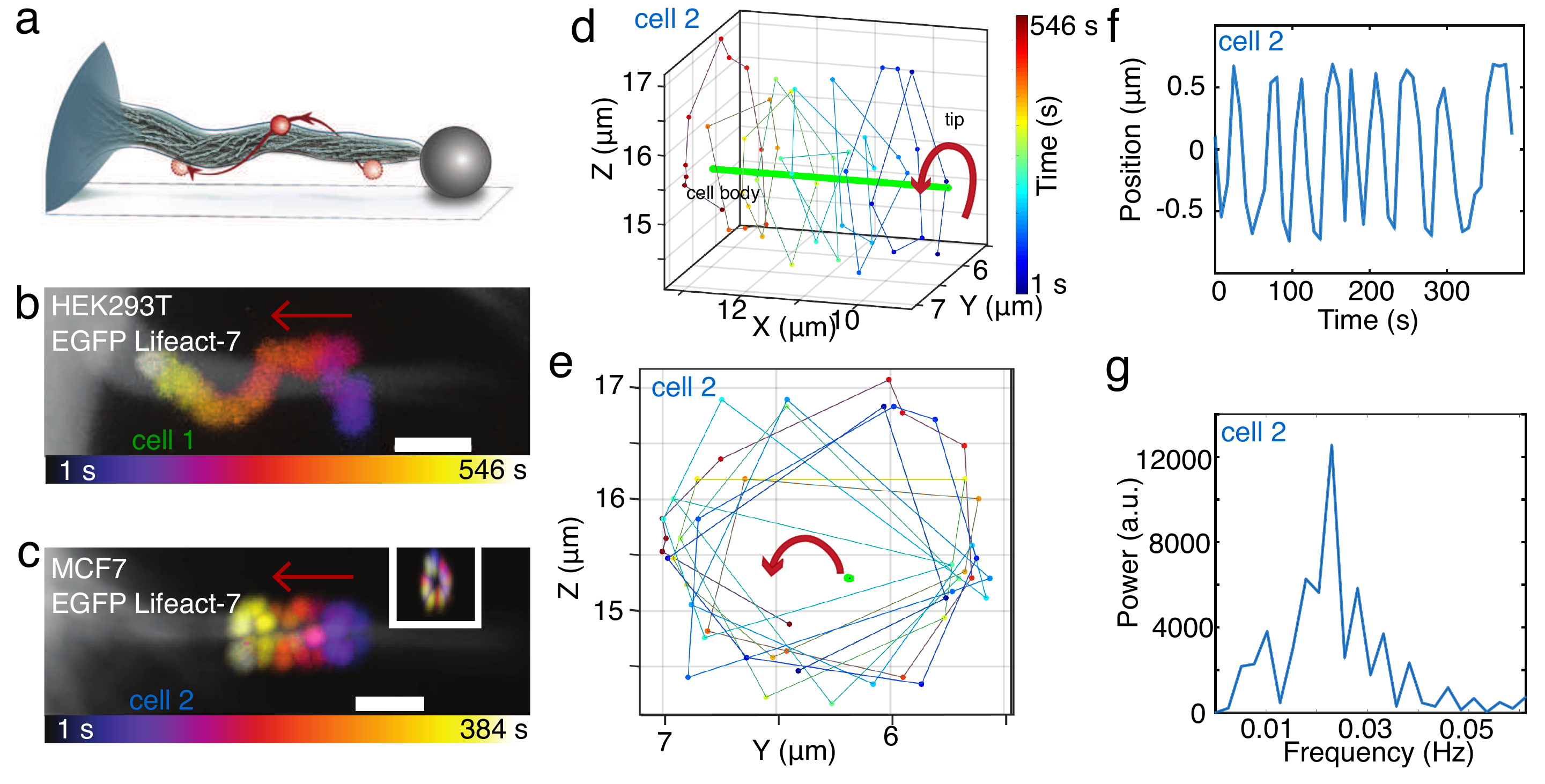}
\caption{{\bf The actin shaft within filopodia performs a spinning motion.}
(a) Assay to visualize internal F-actin rotation in filopodia: A tether of a cell on a glass slide (gray) is extracted from an EGFP Lifeact-7 expressing cell (cyan) using an optically trapped vitronectin coated bead (grey, $d=4.95$~$\mu$m). After tether extraction, the bead is attached to the glass surface of the sample such that the tether is held in place even when the trap is turned off. A tracer bead coated with vitronectin (red, $d=0.99$~$\mu$m) is attached to the tether using the optical trap. The tracer bead binds indirectly to the filamentous actin inside the tether via transmembrane integrins. After the tracer bead is attached, the trap is turned off and confocal $Z$-stacks are acquired at consecutive time points revealing the spinning of the F-actin shaft inside the filopodium. (b and c) Confocal $Z$-projections of a tether from a HEK293T cell (b, cell 1, grey, EGFP Lifeact-7) and a not activated MCF7-p95ErbB2 cell (c, cell 2, grey, EGFP Lifeact-7) at consecutive time points where an attached tracer bead rotates counterclockwise around the actin shaft from tip towards the cell body (color coded from yellow to red). Scale bars are $2$~$\mu$m. Red arrows show direction of motion from tip towards cell body. The inset in (c) shows images of the tracer bead position at all time points in $YZ$-view.
(d and e) $3$D~trajectory of the tracer bead (from blue to yellow, same time scale as in (c) moving counterclockwise around the filopodium (green) towards the cell body in $XYZ$ (d) and $YZ$ (e) view.
(f) Tracer bead position as function of time for cell 2.
(g) The rotation frequency of the tracer bead and thus the actin shaft rotation frequency is found to be $0.022$~Hz, as obtained from the peak of the power spectrum.}
\label{fig:fig2}
\end{figure}

Filopodia-like structures were formed by extraction of membrane tethers from the cell surface by using optical tweezers~\cite{bornschlogl2,kress,leijnse2,pontes1,pontes2,romero}. Depending on the time of observation F-actin can be found to be present inside pulled membrane tethers which subsequently \textcolor{black}{exhibit similar behavior as native} filopodia~\cite{arjonen,bornschlogl1,leijnse2, leijnse1,nurnberg,stylli}. To visualize and quantify any possible actin shaft rotation we use the setup schematically shown in Fig.~\ref{fig:fig2}a. A tether, seen as an artificially extended filopodium, was extracted from an EGFP Lifeact-7 expressing cell (as a marker for F-actin~\cite{riedl}) using an optically trapped vitronectin (VN) coated bead ($d = 4.95~\mu$m). Vitronectin binds to transmembrane integrins which can serve as a link between the bead and the cytoskeletal filamentous actin. Following extraction of the membrane tether the bead was immobilized on the glass surface of the chamber to hold the hereby newly formed filopodium in place. We then used the optical trap to attach another smaller VN coated red labelled bead ($d = 0.99~\mu$m, in the following denoted as Ôtracer beadÕ) to the tether and acquired $Z$-stacks over time using a confocal microscope. The VN coating on the tracer bead ensures that the bead binds to transmembrane integrin proteins which can interact with actin on the cytosolic side of the membrane. As the actin filaments are known to undergo retrograde flow, so will the integrins and the bead attached to the integrins will move along. Hence, this assay should report the movement of the actin shaft inside the tether.\par
Results from this assay are shown in Fig.~\ref{fig:fig2}b and Supplementary Fig.~4 for a bead attached to a tether from a HEK293T cell (`cell 1'). Fig.~\ref{fig:fig2}c-g and Supplementary Movie \textcolor{black}{2a} show bead rotation data for `cell 2', an uninduced MCF7-p95ErbB2 breast carcinoma cell~\cite{egeblad2001,rafn2012}.

As shown in Fig.~\ref{fig:fig2}b-d, \textcolor{black}{Supplementary Fig.~4, and Supplementary Movie 2a}, the tracer bead moves towards the cell body of the HEK293T and the uninduced MCF7-p95ErbB2 cell as expected due to the retrograde flow. The transport of the tracer bead along the filopodium was at the speed of $150$~$\pm$ $131$nm/s, see Supplementary Table~1, and is consistent with retrograde flow of actin within the filopodium. This behavior was also observed for MCF7 cells not expressing the p95ErbB2 receptor, see Supplementary Fig.~5.

We used a custom written MATLAB program to subtract the constant sideway movement of the filopodium and thereby align the axis of the filopodium in the $YZ$- plane (see Supplementary Fig.~6). This revealed that the bead furthermore undergoes a spiralling motion around the filopodium. We isolated the rotation of the bead around the axis of the actin shaft (Fig.~\ref{fig:fig2}d,e) and found it to be clounterclockwise as seen from tip towards the cell body. The rotation frequency of the VN coated tracer bead attached to `cell 2' in Fig.~\ref{fig:fig2}C was $0.002~$Hz as obtained from a power spectral analysis (Fig.~\ref{fig:fig2}f,g). Supplementary Table~1 shows additional tracer bead rotation frequencies measured for beads on filopodia from HEK293T, MCF7, and uninduced MCF7-p95ErbB2 cells.\par

The measured spinning of the actin core is also expected to result in rotational motion of the filopodial tip. Therefore, we next tracked the three dimensional movement of free filopodia and quantified their \textcolor{black}{angular motion}.

\subsection*{The tips of filopodia rotate with a similar angular velocity as the spinning of the actin shaft}

We tracked the motion of F-actin labelled filopodia (via EGFP Lifeact-7 expression) in cells grown on glass slides, Fig.~\ref{fig:fig3}a, and cells embedded in  collagen I gels, Fig.~\ref{fig:fig3}e. The tracking of filopodia was performed using a custom written MATLAB algorithm which allows tracking free tips of filopodia during their growth and shrinkage. The algorithm allowed us to compensate for filopodial drift caused by the migration of the cell and by possible lateral sliding of the filopodium.\par

We acquired confocal $Z$-stacks of cells \textcolor{black}{placed on glass or embedded within } $4$~mg/ml collagen I gels for approximately $5$~min. Fig.~\ref{fig:fig3}a and \ref{fig:fig3}e show an overlay of $Z$-projections of a filopodium from a MCF7 cell \textcolor{black}{on glass} and a KP$^{\text{fl}}$C cell in $4$~mg/ml collagen I, respectively, at different time points. The lateral movement in Fig.~\ref{fig:fig3}a and \ref{fig:fig3}e has been subtracted such that filopodia at all times initiated at the same point, and thus their rotary motion could be tracked in 3 dimensions, see Gabor filtered image of the filopodium in Fig.~\ref{fig:fig3}b shown in $XZ$ (left) and $YZ$ view (right) (Supplementary Fig. 7) .\par

Since the length of a single filopodium varies over time, we chose a common $YZ$ plane close to the tip through which the filopodium crossed at all time points. Fig.~\ref{fig:fig3}c, d and Fig.~\ref{fig:fig3}f, g show that the filopodial tips rotate counterclockwise over time in the $YZ$ plane as seen from the tip towards the cell body for the MCF7 cell on glass and the KP$^{\text{fl}}$C cell in a collagen I matrix, respectively. We obtained the angular velocities (Fig.~\ref{fig:fig3}h and i). The tips of filopodia on the MCF7 and KP$^{\text{fl}}$C cells rotated with a frequency of $0.008$~Hz and $0.004$~Hz, respectively. \textcolor{black}{Fig.~\ref{fig:fig3}i} compares mean angular velocities for HEK293T, MCF7, and uninduced MCF7-p95ErbB2 cells on glass and KP$^{\text{fl}}$C and KPC cells in collagen I matrices. Angular velocities of filopodial tips in $3$D collagen I matrices are clearly lower than of free cells (\textcolor{black}{Fig.~\ref{fig:fig3}i}, inset) indicating that the extracellular fibers confine and slow down the tip movement. The mean values and standard deviations as well as the $p$-values can be found in Supplementary Table~2. \par

\textcolor{black}{Of the $68$ filopodial tips of cells placed on glass or embedded in collagen (see Fig.~\ref{fig:fig3}i and Supplementary Table 2) we found $51$~\% to undergo a counterclockwise rotation (as seen from tip towards the cell body) and $9$~\% were clockwise. In $40$~\% of the cases we could not clearly determine a handedness which is due to a complex behavior of filopodia which includes actin shaft rotation, retrograde flow, buckling and overall cell movement (See example trajectories in Supplementary Fig.~8).}

We also detected filopodia rotation in naive pluripotent mouse embryonic stem cells (HV.5.1 mESCs) grown in 2i~medium, see Supplementary Movie 3a. These cells represent the earliest development and hence could indicate whether filopodia rotation is a general property of cells. These observations together the data from terminally differentiated cells like Hepa 1-6 mouse hepatocytes (Supplementary Movies 3b, c), and invasive cancer cells such as MCF7-p95ErbB2, see Supplementary Table~2, shows that filopodia rotation takes place in both, the earliest and late stages of development.\par

\subsection*{Myosin activity affects filopodia rotation}
\textcolor{black}{To shed light on the active mechanism leading to filopodia rotations we performed mRNA silencing of genes coding for molecular activity within the filopodia. In particular myosin~V and myosin~X motors have been reported to be associated with filopodia formation and activity and have been shown to transport membrane proteins and vesicular content along the filopodium \cite{tamada,Brawley2009,Houdusse2021,Lan2009,Li2017,Jacquemet2019}}.
 
 \textcolor{black}{MCF7 cells were depleted for myosin~Va, Vb or myosin~X by siRNAs which led to significant reduction of the expression levels of these motors in MCF7 cells, see Supplementary Fig.~9. Expression of \textcolor{black}{Lifeact GFP} in MCF7 cells enabled us to track the filopodia in cells, depleted for myosins using siRNAs, and extract the mean angular velocity. MCF7 cells expressing Lifeact GFP, which were depleted for myosin Va by siRNAs, showed a significant reduction in the angular velocity of filopodia when compared to cells transfected with control siRNA (Fig.~\ref{fig:fig3}j). Furthermore, we also measured a reduction in the angular velocity of filopodia in MCF7 cells depleted for myosin~Vb or myosin~X by siRNAs as compared to cells transfected with control siRNA (Fig.~\ref{fig:fig3}j). The chirality of the rotations was also affected by the mRNA silencing of myosin activity (see Supplementary Table~3) leading to higher degree of randomness for the orientation in cells depleted for myosin motors as compared to control cells.} 
 
 \textcolor{black}{Another molecular component important for filopodia function is the formin mDia1. mDia1 has been reported to respond to twist in actin filaments \cite{Kozlov2004,Yu2017, Mizuno2018} and hence could be involved in twisting the actin shaft in filopodia. Expression of mEmerald-mDia1 in MCF7 cells showed that mDia1, besides being uniformly expressed in the cells, also localized to the filopodia of the cells (Supplementary Fig.~10).
 Following depletion of mDia1 in MCF7 cells by siRNA (Supplementary Fig.~11) we did not observe a significant reduction in the angular velocity compared to MCF7 cells transfected with control siRNA (Supplementary Fig.~12).}
 
\textcolor{black}{Overall, these results strongly suggest that the molecular activity from myosin V and myosin X are somehow involved in the rotation of filopodia whereas the actin binding protein mDia1 does not play a role in the observed rotations.} 

\subsection*{Helical buckling and coiling of filopodia}
A clear indicator of axial rotation of actin in filopodia is the presence of helical buckling which arises from over-twisting of the actin shaft. In the following, we therefore focus on filopodia buckling in cells cultured in $3$D~collagen I networks in which dynamic filopodia can build up twist by interacting with the collagen I fibers.
Filopodia from cells grown in $3$D~collagen I gels dynamically explore their 3D environment, but they are found to exhibit a more restricted motion compared to cells grown on glass due to the confinement of the fibers surrounding the cell, as shown in Fig.~\ref{fig:fig4}. Fig.~\ref{fig:fig4}a-d shows a KP$^{\text{fl}}$C cell labelled with mEmerald Lifeact-7, embedded in a $4$~mg/ml collagen I gel, imaged with a confocal microscope. Bending, buckling, and helical coiling of the filopodium become apparent in a color-coded confocal $Z$-stack at $3$ consecutive time points over a time span of $72$~s (Supplementary Movie 4a).\par


\textcolor{black}{The presence of an extracellular matrix could contribute to build-up of twist in the rotating filopodia. Specific and unspecific interactions between filopodia and the surrounding matrix will lead to twist in the rotating structure and further induce buckling in the case that sufficient twist is accumulated.} Filopodia in cells migrating in a $3$D~collagen I network, \textcolor{black}{are expected to experience} external friction at the contact points between filopodium and the fibers \textcolor{black}{\cite{Jacqumet2015}}. Additionally, friction exists between the plasma membrane and the actin shaft mediated by various proteins linking the membrane with the actin. The helical buckles and coils observed in e.g. \textcolor{black}{Fig.~\ref{fig:fig4}a-d}, and \textcolor{black}{Supplementary Fig. 2}, and Movie \textcolor{black}{4a} can hence be a signature of over-twisting of the actin shaft which occurs naturally when rotating filopodia interact with a collagen I network or the membrane. \par

\subsection*{Filopodia buckle and pull at the same time}
To further investigate the nature of these buckles we tested whether such a twist-buckling transition occurs in presence of traction in the filopodium. We therefore measured the traction force when holding the membrane tether by an optical trap, as shown schematically in Fig.~\ref{fig:fig5}a. Compressive buckling from membrane tension, as suggested in Refs.~\cite{derenyi,pronk} is an unlikely mechanism if the filopodium is able to exert a traction force while undergoing buckling. We observed buckling in filopodia, without the presence of external collagen I fibers, in optically trapped filopodia pulled from cells \textcolor{black}{on glass}, see Supplementary Fig.~3 and Supplementary Movie 4b. \par

We extracted tethers from mouse embryonic stem cells and terminally differentiated cells (MCF7) that were transfected to express EGFP Lifeact-7 labelled F-actin, by optical trapping of a vitronectin coated bead \textcolor{black}{($d = 4.95~\mu$m)} (Fig.~\ref{fig:fig5}b, c). After extracting a membrane tether with a trapped bead, small amounts of F-actin were immediately present inside the tether, but over a time course of $\sim 150$~s, the cell recruited more F-actin, as seen in Supplementary Fig.~13a,b for a HV.5.1 mESC, and the actin shaft was observed to extend several micrometers into the membrane tube. Subsequently, the new structure became highly dynamic and behaved like a filopodium (Supplementary Movies 5a,b). The force exerted by the filopodia (Supplementary Fig.~13) was measured over a time span of $60$~s by tracking the beadÕs displacement from the initial center of the trap. \textcolor{black}{To exclude a contribution from the cell motility present in most cell types, we performed parallel tracking of cells and force measurements. As seen in Supplementary Fig.~14 the cell movement is slower than the rapid changes in the force and are therefore uncoupled. The measured force can therefore be ascribed to tension in the filopodium.}\par

To investigate the general traction force delivered by cells existing in different developmental stages we quantified the maximal filopodium traction forces exerted by MCF7, HEK293 cells, and HV.5.1 mESCs cultured in 2i or Serum/LIF (s/L) medium (Fig.~\ref{fig:fig5}d-f). We found that all cell types exclusively exert a traction force in the range~$20$-$80$~pN on the trapped bead and stem cells cultured in 2i or S/L medium were found to exert slightly lower maximal forces. Mean forces and standard deviations and $p$-values can be found in Supplementary Table~4. Time traces of forces from MCF7 cells in Supplementary Fig.~15 show that the extracted tethers display similar activity as expected for filopodia. The high traction forces exceed the typical force of $10$~pN needed for holding a pure membrane tether containing no F-actin \cite{leijnse1,bornschlogl2}. Since buckles can be observed in force generating tethers held by an optical trap we conclude that compressive forces from the membrane are unlikely to be responsible for the observed buckling and coiling of filopodia, but instead our data suggest that the actin structure exhibits an internal twist generating mechanism. 



\subsection*{Twist deformations are a generic non-equilibrium feature of confined actomyosin complexes}

Our experimental observations demonstrate that the  complex dynamics of filopodia including helical rotation, buckling, and tip movement are induced by the twisting motion of actin filaments inside the filopodia. Furthermore, the emergence of such twisting motion in a variety of cell types and in both early and late stages of development points to a possible generic mechanism for the formation of twist in actin/myosin complexes within the filopodial cell membrane. To understand the underlying mechanism of twist generation, we next used a three-dimensional active gel model to study the dynamics of actin/myosin complexes confined within a geometrical constraint. \textcolor{black}{The existence of molecular activity in filopodia is well established through the presence of actin reorganization proteins \cite{medalia2007organization,RN242,shekhar2015formin} and molecular motors like myosin~V and myosin~X \cite{tamada,Brawley2009,Houdusse2021,Lan2009,Li2017}.} The choice of model was motivated by the generality of this class of continuum models which is due to the fact that only local conservation laws, interaction between the systemsÕ constituents, and perpetual injection of energy at the smallest length-scale are assumed. Furthermore, active gel equations have proven successful in describing several aspects of the physics of actin/myosin networks including actomyosin dynamics at the cell cortex~\cite{naganathan14,julicher18}, actomyosin induced cell motility~\cite{tjhung17,banerjee20}, actin retrograde flows~\cite{julicher07,Prost15}, and topological characteristics of actin filaments~\cite{zhang18,kumar18}. Within this framework the dynamics of actin/myosin complexes are expressed through a continuum representation of their minimal degrees of freedom including the orientation and velocity fields. The orientation field is represented by a tensor order parameter ${\mathbf{Q}}=3q/2\times(\vec{n}\vec{n}^{\mbox{T}}-{\mathbf{I}}/3)$, where $q$ is the magnitude of the orientational order and $\vec{n}$ is the director, representing the coarse-grained orientation of the actin filaments~\cite{Marchetti2013,Doostmohammadi2018}. The dynamics of the orientation tensor $\mathbf{Q}$ follows Beris-Edwards equations~\cite{BerisBook}, describing the alignment to flow and relaxation dynamics due to the filament elasticity $K$. The orientation dynamics is coupled to the velocity field governed by a generalized Stokes equation that accounts for the active stress generation due to force dipoles associated with actin tread-milling, described as $\mathbf{\Pi}^{\mbox{active}}=-\zeta\mathbf{Q}$, where $\zeta$ denotes the strengths of the active stress generation (see {\it Supplementary Information} for the details of the governing equations and the mapping of parameters to physical units). 

In order to emulate the dynamics of actin filaments inside the filopodia, we consider a simplified setup of an active gel confined inside a three-dimensional channel with a square cross-section of size $h$. In two-dimensions it is shown that increasing the confinement size above a threshold results in a spontaneous shear flow inside the confinement due to bend-splay deformations~\cite{Voituriez2005,Duclos2018}. Interestingly, we find that here increasing the size of the three-dimensional confinement beyond a threshold results in the spontaneous flow generation accompanied by the emergence of twist deformations (Fig.~\ref{fig:SIM}a,b). To characterize the amount of the twist in the system we measure the average twist deformations across the channel $\mathcal{T}=\langle|\vec{n}\cdot\left(\vec{\nabla}\times\vec{n}\right)|\rangle_{\vec{x},t}$, where $\langle\rangle_{\vec{x},t}$ denotes averaging over both space and time. As evident from Fig.~\ref{fig:SIM}c increasing the confinement size above a threshold results in increasing amount of twist in the active gel. The amount of the twist generation and the threshold are further controlled by the activity of the filaments $\zeta$ and their orientational elasticity $K$. For a fixed channel size $h$ increasing activity triggers a hydrodynamic instability~\cite{Simha2002,Voituriez2005} that results in spontaneous flow generation and twist deformations (Fig.~\ref{fig:SIM}d) consistent with the experimental observations in Fig.~\ref{fig:fig4} and Supplementary Fig.~16. The hydrodynamic instability and the creation of the spontaneous flows are suppressed by the filament elasticity $K$. As such increasing the filament elasticity reduces the amount of twist in the system up to a threshold value, where any twist generation is completely suppressed (Fig.~\ref{fig:SIM}e).

The interplay between the activity, elasticity, and the confinement size can be best understood in terms of a dimensionless parameter $\mathcal{A}=h\times\sqrt{\zeta/K}$ which describes the competing effect of two length scales: the confinement size $h$ and the length scale set by combined effects of activity and elasticity $\sqrt{K/\zeta}$. Indeed plotting the amount of twist as a function of the dimensionless number $\mathcal{A}$ results in the collapse of the data corresponding to varying activity, elasticity, and confinement size (Fig.~\ref{fig:SIM}f), indicating that $\sqrt{K/\zeta}$ is the relevant activity-induced length scale: increasing the activity enhances active stress generation and orientational deformations, while such deformations are accommodated by the elasticity. As a result, larger activity (or smaller elasticity) leads to the emergence of deformations with smaller length scales. On the other hand, suppression of the activity is expected to increase the deformation length scale. When this length scale becomes larger than the confinement size, all deformations are suppressed and no twist is expected in the actin filaments. 

\subsection*{Chirality of the twist} 
It is important to note that the emergence of twist happens without having any explicit chirality in the equations of motion and is due to a hydrodynamic instability that is induced by the activity of the confined actomyosin complexes. Since this hydrodynamic instability breaks the mirror-symmetry, it will choose both clockwise or counterclockwise directions of rotation with a similar probability. From a biological perspective there could be several contributors for biasing the orientation of the twist. The molecular motors  myosin~V \cite{ali2002} and myosin~X \cite{sun2010} have been shown to walk along a spiral path on actin bundles found in filopodia. These motors walk toward the tip in a counterclockwise orientation which would contribute to a torque in the opposite direction. Our experiments suggest that the rotation of filopodia is biased towards the opposite orientation of the path formed by these myosin motors which could suggest that motors increase the probability for the measured orientation. To capture this, experimentally measured bias, in our active gel dynamics we refine our model to include active stress originating from torque dipoles. 

The effect of torque dipoles is accounted for through additional contributions to the active stress $\Pi_{\alpha\beta}^{\mbox{active torque}}=-\zeta'\epsilon_{\alpha\beta\gamma}\partial_{\mu} Q_{\gamma\mu}$, where $\epsilon_{ijk}$ is the Levi-Chevita operator and $\zeta'$ controls the strength of the torque dipole~\cite{Julicher12,naganathan14}. It is easy to see that this term already breaks the mirror-symmetry, such that positive (negative) $\zeta'$ exerts clockwise (counterclockwise) torque. No experimental measure of the relative strength of force and torque dipoles are available, however, dimensional analyses suggests that the ratio of active stress coefficients $\zeta'/\zeta\sim 0.1$ indicating that in general for actomyosin complexes the contributions of torque dipoles to the dynamics are small compared to the force dipoles (see Supplementary Information for the estimate of the torque dipole coefficient). Furthermore, the emergence of small, but finite number of clockwise rotations in the experiments, show that the hydrodynamic instability is the controlling mechanism for the mirror symmetry breaking. Nevertheless, even small values of this chiral terms will bias the mirror-symmetry breaking in the direction of the torque dipole. Indeed our representative simulations show how changing the sign of $\zeta'$ results in the change in the handedness of the filaments rotation around the axis of the channel (see Supplementary Fig.~17).

\section*{Discussion}
Our results show that axial twisting and filopodia rotation is generic in cells and is detected in naive pluripotent stem cells and in terminally differentiated cells. The spinning of the actin shaft leads to a rich variety of physical phenomena such as tip movement, helical buckling, and traction force generation and this allows the cell to explore the $3$D extracellular environment while still being able to exert a pulling force as summarized in Fig.~\ref{fig:fig7}.\par

The rotational behavior of filopodia has been largely unexplored in literature. However, the periodic sweeping motion of macrophage filopodia was found to be $1.2$~rad/s ($0.191$~Hz) as measured in two dimensions~\cite{zidovska}. Growth cone filopodia were also tracked in two dimensions orthogonal to the axis of the filopodium and the frequency was found to be $\sim 0.016$~Hz~\cite{tamada}. In Ref.~\cite{leijnse2} the actin structure of filopodia was imaged and small buckles were observed to rotate around the actin shaft, thus strongly indicating that actin has the ability to spin within the filopodium.\par

We explicitly performed $3$D tracking of the whole filopodium and its tip, see \textcolor{black}{Fig.~\ref{fig:fig7}c-d}, and made complimentary measurements of the spinning of the actin shaft within the filopodium, see \textcolor{black}{Fig.~\ref{fig:fig7}f-h}. \textcolor{black}{Beads were added to the side of the filopodium by an optical trap and subsequently tracked in 3D. Vitronectin on the beads allows for binding directly to intergrin which can couple the bead to the internal actin structure. We note that beads can also bind non-specifically to plasma membranes \cite{Seo2016} or to filopodia and still be coupled to the actin structure, as shown in Ref.\cite{Kohler2015} and in Supplementary Movie 2b. A clear signature that the bead is connected to the actin is translocation of the bead towards the cell body which is caused by retrograde flow within the filopodium.} The results from these experiments showed that the actin shaft has the ability to spin with a similar frequency as the circular movement of the whole filopodium. These results strongly suggest that the spinning of the actin together with filopodia bending and growth, is responsible for the $3$D motion and buckling of filopodia.\par

When a filopodium is free, we find that the tip rotates \textcolor{black}{with a mean angular velocity of $0.4\pm0.4$~deg/s. In the presence of friction, arising from cells being embedded in a $3$D matrix, we observe that rotation slows down and filopodia rotate with a mean angular velocity of $0.1\pm 0.1$~deg/s} The external resistance to rotation experienced by filopodia leads to accumulation of twist and hence to helical buckling of the filopodium (Fig.~\ref{fig:fig7}-d). This mechanism can be explained by twisting a rubber cable with one hand while holding the other end tight. The cable will accumulate twist, start to buckle and coil into an helical shape as shown in Supplementary Fig.~16. During buckling and coiling such a structure will shorten and therefore generate a pulling force. \par

When actin is not specifically linked to an external structure which can prevent spinning of the actin, buckling can still occur if friction exists between the membrane and the actin structure. In the presence of friction, for example from transmembrane proteins such as integrins and peripherally binding proteins like I-BARs~\cite{hu1}, torsional twist can accumulate in the actin shaft and cause buckling~\cite{leijnse2,wada}. Buckles form when twist is released and converted into bending energy within the actin shaft. This leads to an apparent shortening of the filopodium as observed when holding the filopodium by an optical tweezer~\cite{leijnse2} and thus contributes to a traction force in addition to the force arising from the retrograde flow of actin~\cite{bornschlogl2} \textcolor{black}{(Fig.~\ref{fig:fig7}i-k)}.\par

\textcolor{black}{We found the filopodia to predominantly rotate in a counterclockwise orientation, as seen from the tip towards the cell body. Of all resolvable rotations (N = 68, Fig.~\ref{fig:fig3}i, Supplementary Fig.~8 and Supplementary Table~2) we found $51$~\% of the filopodia to rotate in a counterclockwise orientation while $9$~\% rotated in a clockwise orientation as seen from the tip towards the cell. The orientation of $40$~\% of the filopodia rotations could not clearly be resolved due to convoluted lateral movements, buckling and rotations. Furthermore, in the control experiments in Fig.~\ref{fig:fig3}j and Supplementary Table 3 we observed a bias towards counterclockwise tip rotation with $49$~\% exhibiting counterclockwise rotation, $15$~\% rotated clockwise while $36$~\% of rotation directions could not be resolved for a total of N = 92 filopodia.} Interestingly, a similar sweeping motion  was also measured for the movement of growth cone filopodia in Ref.~\cite{tamada} which could indicate that filopodia are generically prone to exhibit counterclockwise rotation when observed from the perspective of the tip. However, to properly resolve the orientation of the spinning actin shaft it is necessary to perform experiments as in Fig.~\ref{fig:fig2} where the actual spinning of the actin shaft can be isolated from the lateral movement of the filopodia.\par

Twist in actin bundles emerges naturally when active filaments are confined in channels with dimensions similar to a filopodia tube as shown by active matter simulations, \textcolor{black}{(Fig.~\ref{fig:SIM} and Fig.~\ref{fig:fig7})}. \textcolor{black}{Our model assumes active force generation along the actin shaft which could arise from walking motion of myosin motor proteins along actin filaments. The presence of myosin~V and myosin~X in filopodia of cancer cells and other cell types is well established \cite{tamada,Brawley2009,Houdusse2021,Lan2009,Li2017} and these motors can produce force and torque dipoles as shown schematically in Fig.~\ref{fig:fig7}a}. Both of these motors experience a drag while walking along the actin bundle: myosin~V from dragging vesicle encapsulated cargo while myosin~X causes drag from transporting membrane embedded proteins along the viscous plasma membrane as it walks towards the tip of the filopodium. Indeed we measured reduction in filopodia rotations upon mRNA silencing of myosin~V or myosin~X which supports the force generating ability of these motors on the actin structure. The induced chirality of the rotations is random, but can be biased by active torque dipoles in the system. Such dipoles indeed exist in natural filopodia in the form of myosins exhibiting chiral motion and interfilament stepping along actin bundles as demonstrated for the filopodia associated motors myosin~V \cite{ali2002} and myosin~X \cite{sun2010}. Both of these motors walk towards the tip in a filopodium and spiral around the actin in a clockwise orientation as seen from the tip thus inducing counterclockwise rotations, as seen from the perspective of the tip. Our experimental data indeed show a preference for counterclockwise orientation of the twist \textcolor{black}{, see Supplementary Table 2}, \textcolor{black}{whereas mRNA silencing of myosin~V and myosin~X resulted in increased randomness for the orientation (Supplementary Table~3) thus supporting the idea that torque-generation by these motors induces a bias towards a specific chirality of the rotations.}  The presence of both twist-orientations supports the theoretical predictions of a hydrodynamic instability being the origin of the twist.

To exclude possible artifacts arising from the transiently expressed fluorescent actin we also imaged cells which had the actin labelled with a membrane permeable probe which binds to F-actin in living cells (SiR actin). These cells showed similar filopodial activity as we observed in cells transfected with EGFP Lifeact-7, see Supplementary Movies S3b,c.\par

The observation of similar filopodial dynamics in early stem cells confirms the generic nature of filopodial dynamics. The state of stem cells used here resemble the inner cell mass (ICM) cells, but their state can be controlled by different culture media. Cells cultured in serum-free medium supplemented with leukemia inhibitor factor (LIF) and small molecule inhibitors GSK3 and MEK (2i~medium), genetically represent cells from the ICM of mouse blastocyst corresponding to $3.5$ days post fertilization and are highly pluripotent~\cite{sim,morgani}. Cells grown in Serum-LIF (S/L) medium are in a primed state towards epiblast~\cite{morgani}. The colonies of cells grown in these two media exhibit remarkably different morphologies; cells cultured in S/L media are more spread out (Supplementary Fig.~13d) and form a monolayer of cells while 2i~cultured cells grow together into an embryonic body (Supplementary Fig.~13c). The general role of filopodia in early development is less clear. Their presence has been detected in both the process of embryonic compaction, where they were shown to sustain tension, which was concluded form images, but the force was not quantified~\cite{fierrogonzalez,pillarisetti}. Our quantitative data show that naive pluripotent embryonic stem cells are able to deliver a traction force on the order of $10$~pN for each filopodium. However, whether the spinning motion of actin does occur during compaction, where the filopodia are penetrating other cells, remains an open question. Filopodia in mesenchymal stem cells have been shown to function as rails facilitating transport of morphogens between cells~\cite{sanders}. Such tunnelling tethers or cytonemes have been observed in many cell types and we have also detected rotation of such structures in Hepa 1-6 mouse hepatocytes (Supplementary Movies 3b,c), however, the functional role of the rotation remains elusive in these structures.

\section*{Conclusion}
Filopodia are extremely versatile surface structures with ability to explore the $3$D extracellular space by simultaneous growth, bending, and rotary motion. Together, our experiments and theoretical modeling presents new evidence of a general rotary mechanism observed in filopodia. Remarkably, we report this rotary mechanisms for a wide selection of different cell types ranging from na\''i ve stem cells to terminally differentiated cancer cells, indicating that the observed phenomenon must be generic within various cell types. We find that the actin shaft inside filopodia generally spins and is responsible for the circular motion of filopodia which allows them to explore their $3$D extracellular environment. Using a hydrodynamic model of actin filaments we show that the spinning can originate from symmetry breaking and subsequent twist generation by active filaments under a confinement with similar dimensions as the filopodia tube. \textcolor{black}{Filopodia rotations were reduced by silencing either myosin~V or myosin~X and the chirality of the rotations became less biased. The activity from these motors, which walk in a helical path around actin bundles, is therefore responsible for spinning and twisting the actin core of filopodia.} Signatures of twist are observed experimentally as helical buckles and coils in a variety of cell lines and can give rise to shortening of the actin shaft and therefore generation of a traction force. Additionally, the functional implication of twist generation in filopodia for cellular processes could be to allow the cell to explore their 3D extracellular environment, to navigate through dense networks of the ECM and to assist in chemical sensing.

\section*{Acknowledgements}
PMB and NL acknowledge support from the Danish Council for Independent Research, Natural Sciences (DFF-4181-00196) and LBO acknowledges support from the Danish National Research Foundation (DNRF116). MRA and JN acknowledge support from the Novo Nordisk Foundation (NNF18OC0034936). AD acknowledges support from the Novo Nordisk Foundation (grant No. NNF18SA0035142), Villum Fonden (Grant no. 29476), Danish Council for Independent Research, Natural Sciences (DFF-117155-1001), and funding from the European UnionÕs Horizon 2020 research and innovation program under the MarieSklodowska-Curie grant agreement No. 847523 (INTERACTIONS). JTE acknowledges support from the Novo Nordisk Foundation (Hallas M¿ller Stipend). We thank Joshua Brickmann for providing us with embryonic stem cells, Anne Benedicte Mengel Pers for providing hepatocytes and Agnieszka Kawska for help with schematic illustrations.

\subsection*{Author contributions}
P.M.B. initiated and supervised the study. N.L., performed optical trapping and imaging experiments and performed all statistical analysis, Y.F.B. performed all tracking of filopodia, S.L.S., M.R.A., carried out silencing of cells, M.R.A., Y.F.B., N.L. conducted imaging of silenced cells and control cells, B.V. and N.L. performed tether pulling experiments, L.W. and N.L. carried out imaging of cells embedded within collagen, S.S. purified plasmids for imaging of actin, L.B.O. supervised optical trapping and provided optical tweezers setup, A.D. carried out active matter simulations, N.L., P.M.B., A.D. wrote the first version of the manuscript. Conceptualization of research: P.M.B, A.D., L.B.O., J.N., J.T.E., all authors read and approved the manuscript for publication.

\subsection*{Competing interests}
The authors declare no competing interests.

\section*{Data availability}
The data that support the findings of this study are available from the corresponding author upon reasonable request.
\newpage

\section*{Figures}

\clearpage

\clearpage

\begin{figure}
    \captionsetup{labelformat=empty}
    \centering
 \includegraphics[scale=0.5]{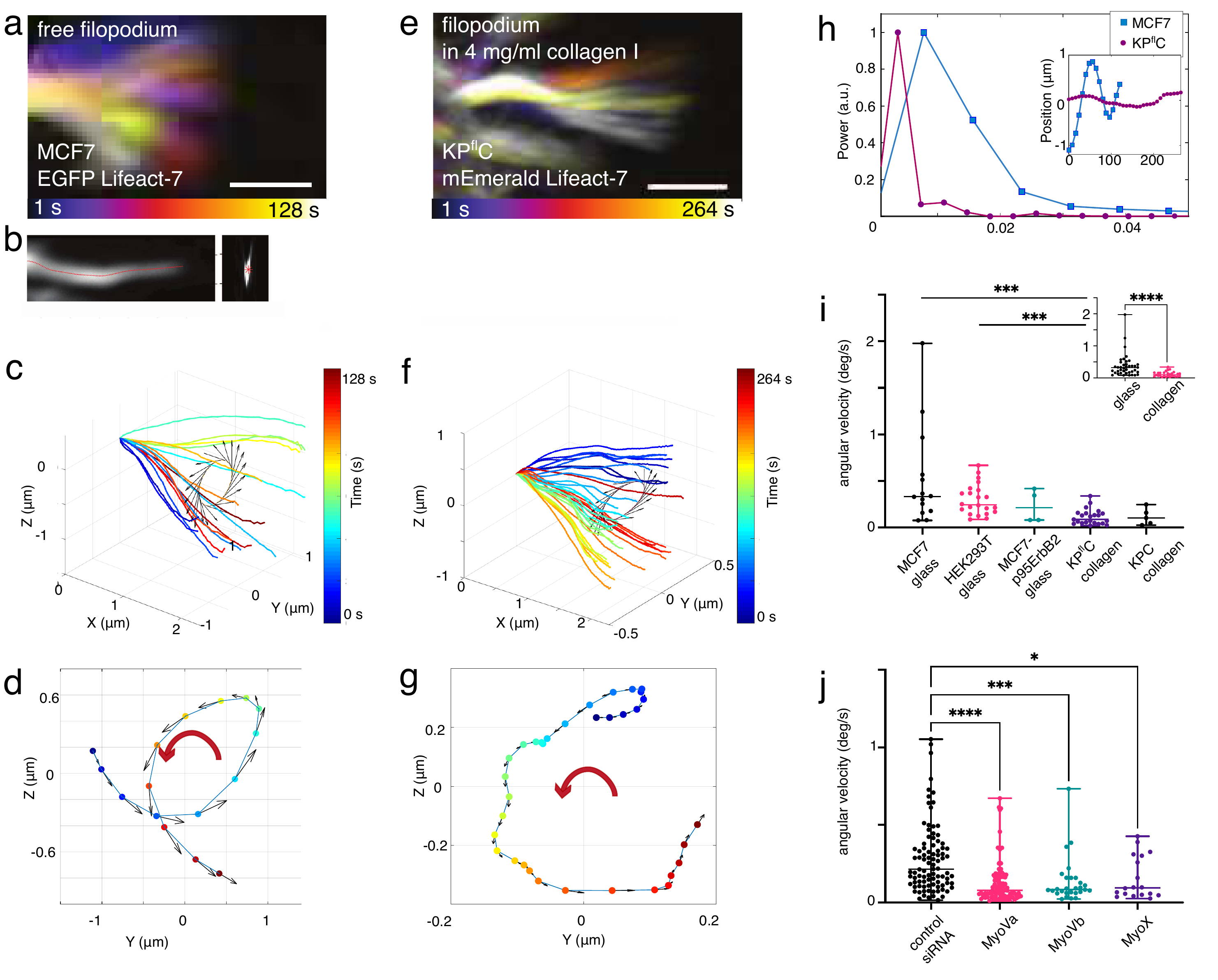}
    \caption{}
\end{figure}
\clearpage
\begin{figure}
    \ContinuedFloat
\caption{{\bf The tips of free filopodia of cells \textcolor{black}{on glass} and filopodia of cells grown in a collagen I matrix undergo a rotary motion with different angular velocities.}
(a) Gabor-filtered $Z$-projection of the filopodial tip movement of a EGFP Lifeact-7 labelled MCF7 cell at $16$~consecutive time points for a total time of $128$~s. Scale bar is $1$~$\mu$m.
(b) $XZ$ (left) and $YZ$ (right) view of a single $Z$-projection showing the Gabor filtered data (grey) overlayed with the traced filopodium (red).
(c and d) Result from tip tracking of the filopodium from (a) in $XYZ$ (c) and $YZ$ view (d). Tip tracking over time reveals a counterclockwise rotation of the whole filopodial tip, seen from tip towards the cell body, as indicated by the red arrow.
(e) Gabor-filtered $Z$-projection of filopodial tip movement of an EGFP Lifeact-7 labelled KP$^{\text{fl}}$C cell grown in $4$~mg/ml collagen I at $16$~consecutive time points for a total time of 264~s. Scale bar is $1$~$\mu$m.
(f and g) Result from tip tracking of the filopodium shown in (e) in $XYZ$ (f) and $YZ$ view (g). Tip tracking over time reveals a counterclockwise rotation of the whole filopodial tip, seen from tip towards the cell body, as indicated by the red arrow.
\textcolor{black}{(h) The rotation frequency of the filopodial tip for the MCF7 cell from (a) is $0.008$~Hz and the KP$^{\text{fl}}$C cell in $4$~mg/ml collagen I from (e) is $0.004$~Hz. Inset: Filopodial tip position as function of time.}
\textcolor{black}{(i) Angular velocities of filopodial tips from cells on glass (MCF7, (N filopodia = 14), HEK293T (N = 22), MCF7-p95ErbB2 (N = 4)) and in collagen I (KP$^{\text{fl}}$C (N = 23), KPC (N = 5)). Kruskal-Wallis statistical test was used with significance set at $p$ < 0.05. Inset: Pooled angular velocities of filopodia from cells grown on glass (N filopodia = 40) and in collagen I (N = 28). Mann-Whitney statistical test was used. Scatter plots show the median and the whiskers extend from minimum to the maximum values. $p$ values are listed in Supplementary Table 2. 
(j) Angular velocities of filopodia after silencing myosins Va (N (filopodia)=88), Vb (N=28), and myosin X (N=18), respectively, were compared to cells exposed to control siRNA (N=92). Kruskal-Wallis statistical test was used with significance set at $p$<0.05. $p$<0.0001, $p$=0.0003, $p$=0.0171, for control siRNA vs MyoVa, Vb, and X, respectively. Scatter plot shows the median and the whiskers extend from minimum to the maximum values.}}
\label{fig:fig3}
\end{figure}
\clearpage

\begin{figure}[h]
 \centering
 \includegraphics[scale=0.5]{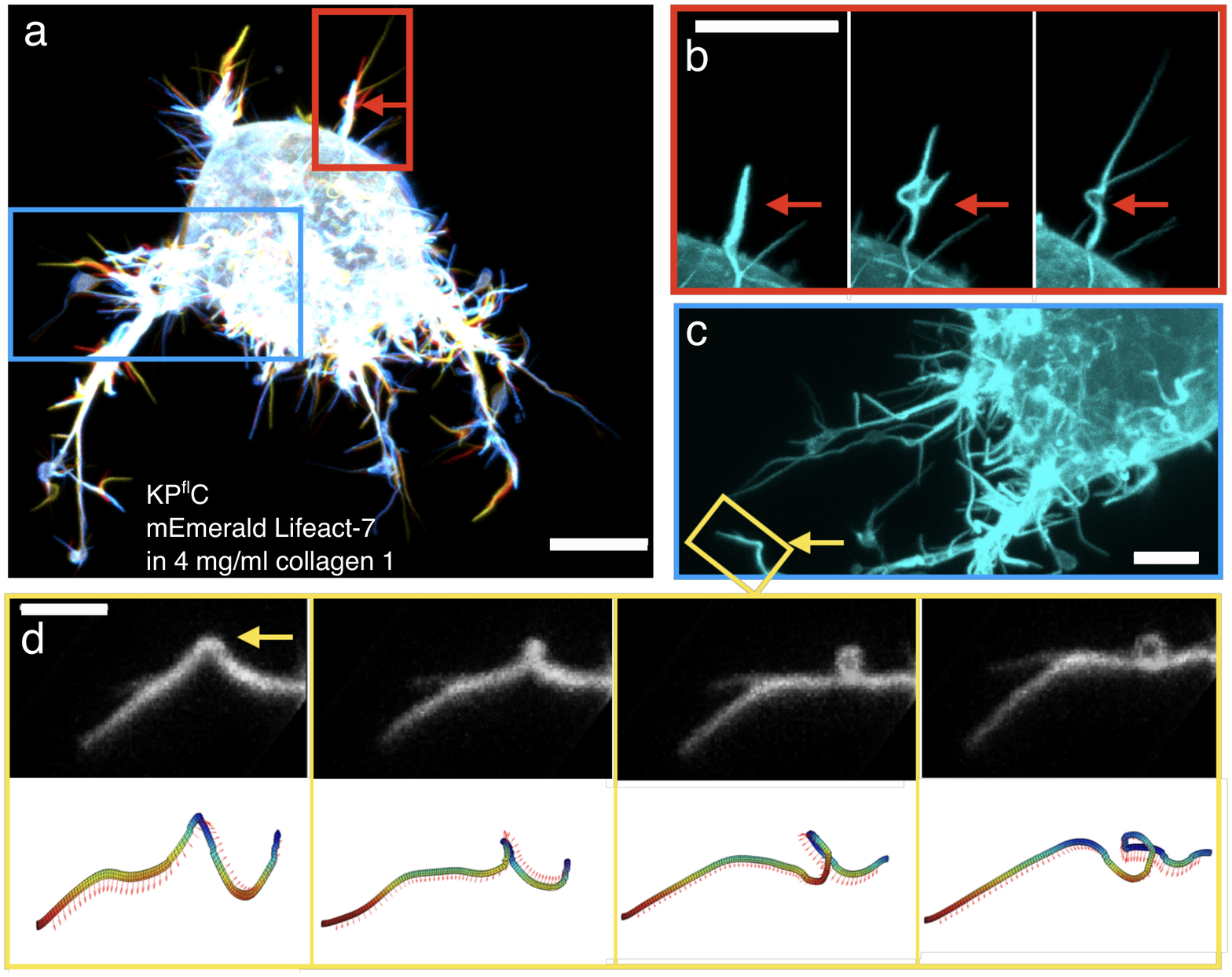}
\caption{{\bf Filopodia of cells embedded in a collagen I matrix undergo helical buckling as a result of rotation and twist accumulation upon contact with the matrix.}
(a) Overlay of $3$ confocal $Z$-projections of a KP$^{\text{fl}}$C cell (mEmerald Lifeact-7) embedded in $4$~mg/ml collagen I at consecutive time points (color coded over $72$~s). The red arrow highlights a region where a buckle forms. Scale bar is $10$~$\mu$m.
(b) Zoom-in on the red rectangular region from (a): Individual images of the $3$ time points show buckle formation. Scale bar is $10$~$\mu$m, frame interval is $24$~s.
(c) Zoom-in on the blue rectangular region from (a) (rotated). The yellow rectangle highlights a bending filopodium. Scale bar is $5$~$\mu$m.
(d) Top: Gabor-filtered images of a the filopodium from the yellow region in (c) at $4$ consecutive time points (frame interval is $19$~s) of a KP$^{\text{fl}}$C cell in $4$ mg/ml collagen I. Scale bar is $2.5$~$\mu$m. Bottom: $3$D tracks of the filopodium show bending and coiling. Red arrows indicate the binormal vectors along the filopodium.}
\label{fig:fig4}
\end{figure}
\clearpage

\begin{figure}[h]
 \centering
 \includegraphics[scale=0.5]{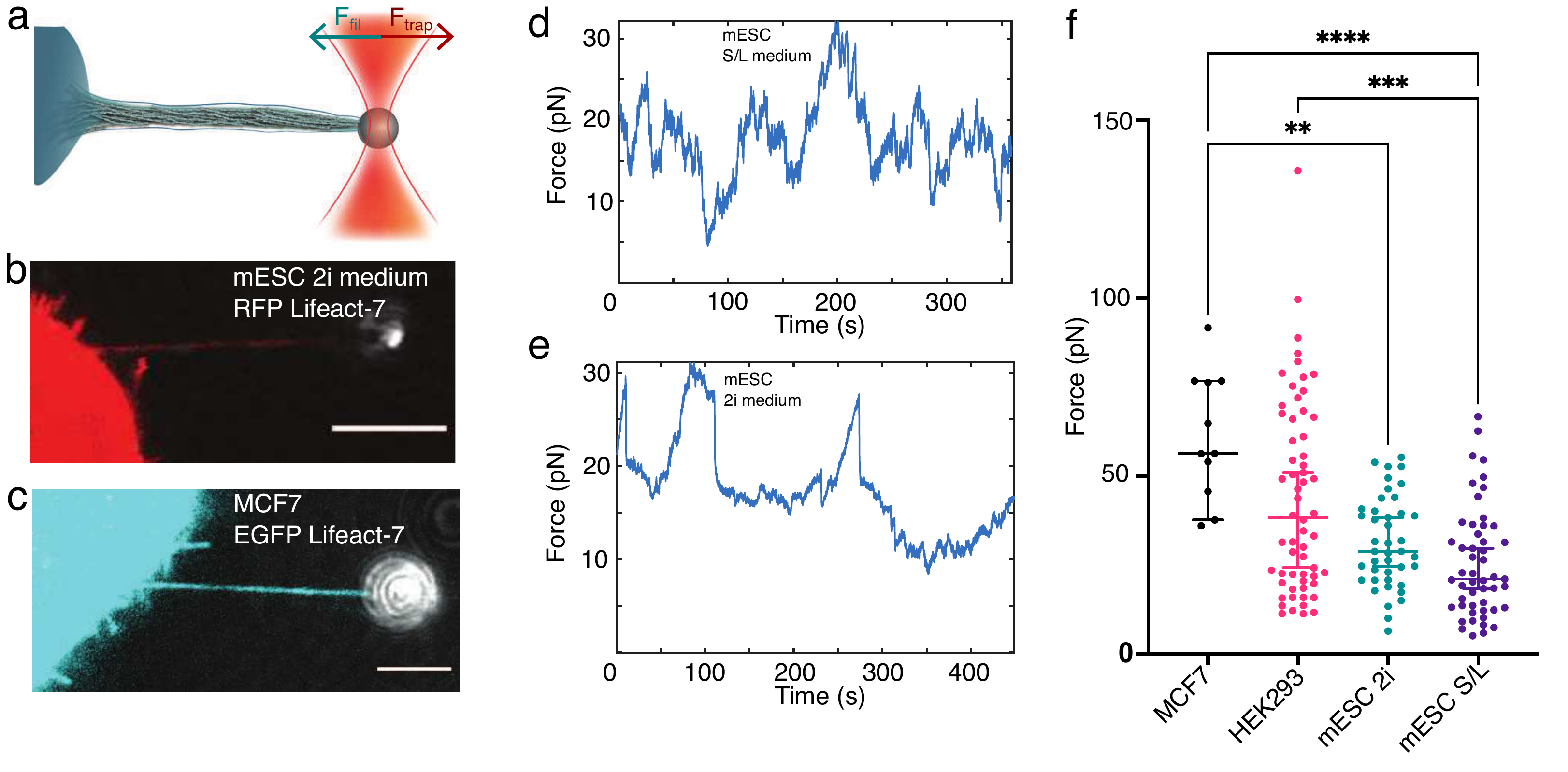}
\caption{{\bf Tethers pulled from different cell types-- from naive pluripotent stem cells to terminally differentiated cells-- are highly dynamic and generate significant traction forces.} 
(a) Schematics of the optical tweezers assay to measure the dynamic force exerted by a tether. An optically trapped VN coated bead ($d=4.95$~$\mu$m) is used to extract and hold a membrane tether at a given length with a holding force $F_{trap}$. (b) Confocal image of a filopodium extracted from a HV.5.1 mESC (red, RFP Lifeact-7) grown in 2i~medium. Scale bar is $5$~$\mu$m. 
(c) Confocal image of a filopodium extracted from a MCF7 cell (cyan, EGFP Lifeact-7). Scale bar is $5$~$\mu$m.
(d) and (e) Force curves for tethers pulled from HV.5.1 mESCs grown in 2i (d) and in S/L (e) medium.
(f) Maximum tether holding forces measured for MCF7 (N = 11, mean $\pm$ std = ($61.1 \pm 17.9$)~pN), HEK293 (N = 60, ($43.9 \pm 27.1$)~pN), HV.5.1 mESCs cultured in 2i (N = 42, ($31.3 \pm 12.7$)~pN) and S/L medium (N = 51, ($25.6 \pm 15.4$)~pN). \textcolor{black}{Kruskal-Wallis test with significance set at $p$ < 0.05. Scatter plot shows the median and the whiskers extend from minimum to the maximum values.} $p$-values are shown in Supplementary Table~4.}

\label{fig:fig5}
\end{figure}
\clearpage

\begin{figure}[h]
 \centering
 \includegraphics[scale=0.3]{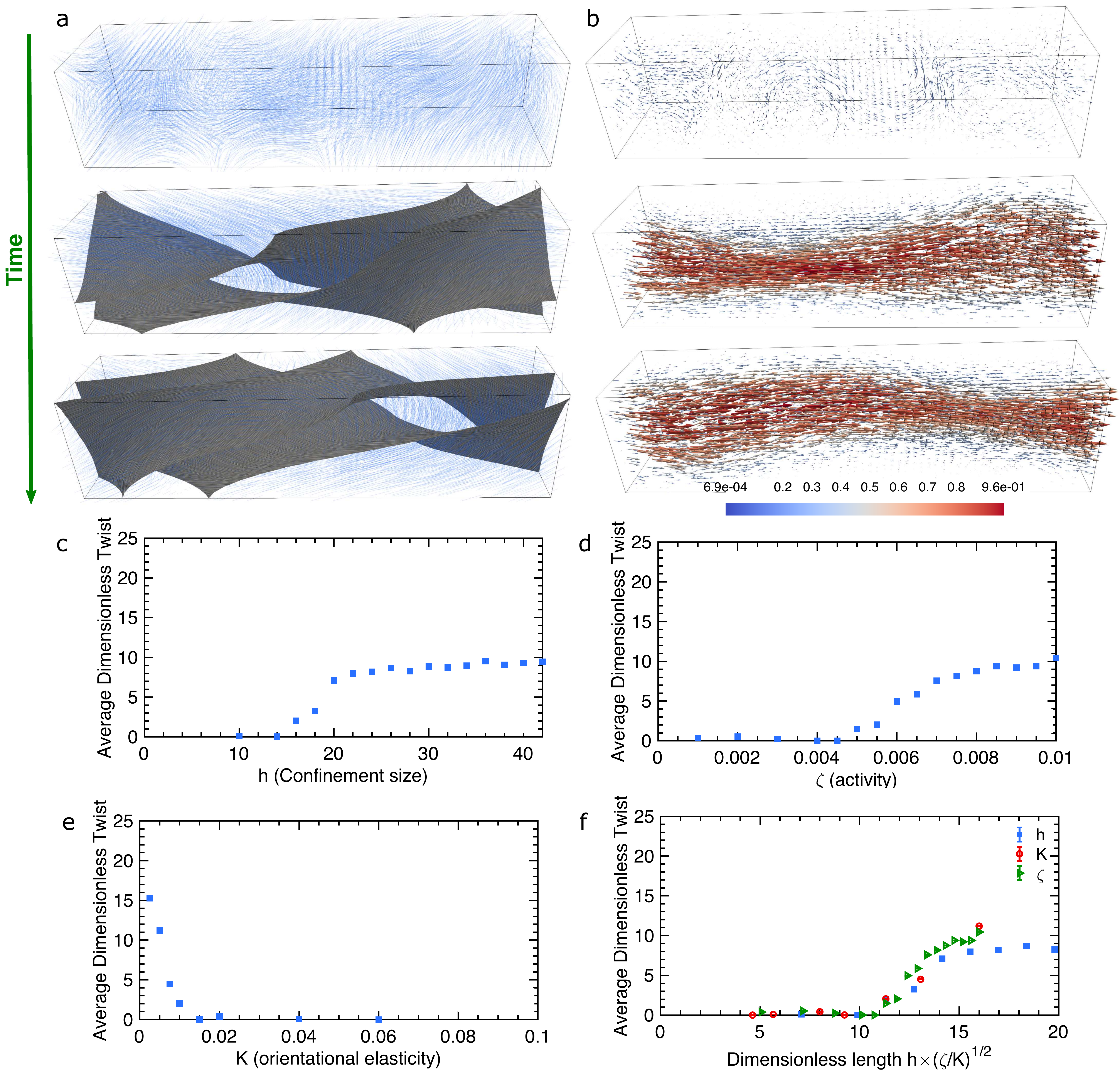}
\caption{{\bf Actin shaft twisting is a generic phenomenon caused by the activity of actomyosin complexes inside the confining filopodial cell membrane.}
(a,b) Temporal evolution of (a) the orientation and (b) velocity fields of the active gel representing actin filaments/myosin motor mixtures. In a the director field associated with the coarse-grained orientation of actin filaments $\vec{n}$ is shown by blue solid lines, and is overlaid with the isocontours of twist deformations $\vec{n}\cdot(\vec{\nabla}\times\vec{n})$. In b the colormap indicates the magnitude of the velocities normalized by the maximum velocity. (c,d,e) Dependence of the average twist on (c) the confinement size, (d) activity, and (e) elasticity. The average amount of twist is non-dimensionalized by the channel length. (f) Average twist as a function of the dimensionless length, for varying confinement sizes, activities, and elasticities, showing the collapse of the data.}
\label{fig:SIM}
\end{figure}
\clearpage

\begin{figure}[h]
 \centering
\includegraphics[scale=0.7]{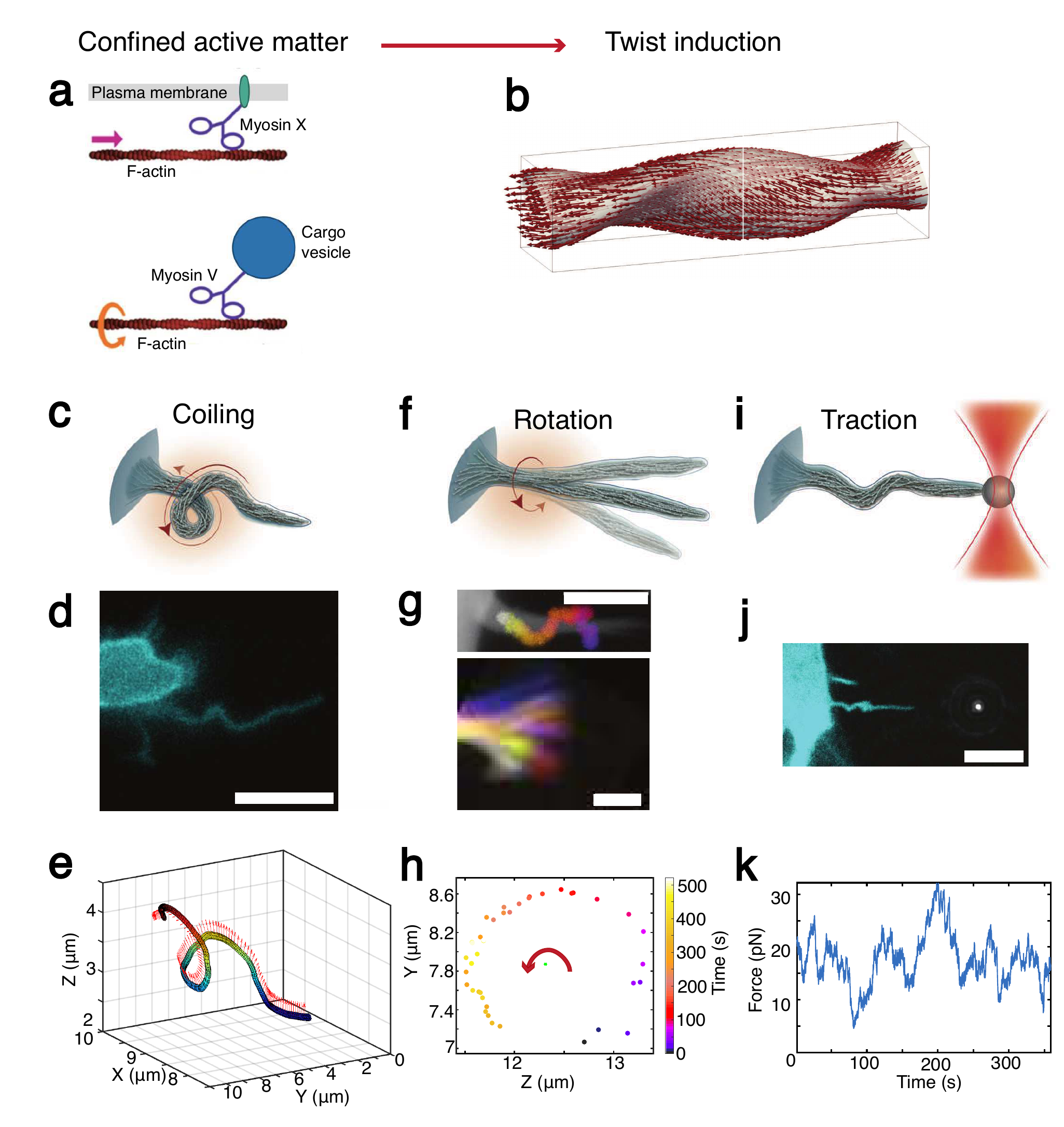}
\caption{}
\label{fig:fig7}
\end{figure}
{\bf Twisting of the active filamentous actin core of filopodia is responsible for coiling, rotation, and traction by filopodia.} \textcolor{black}{(a) Schematic of how walking of myosin motor on actin bundles with inherent helical structures results in generation of force and torque dipoles. Both myosin X and V walk in a helical motion around actin bundles and have the ability to make interfilament steps \cite{ali2002,sun2010}. The drag experienced from the viscous membrane (myosin X) and also from the attached cargo (myosin V) results in local forces applied to the actin structure. Molecular force dipoles originating from other myosins present in the cell cortex have been described previously \cite{naganathan14}.} (b) Filopodia twist is induced by active filaments under confinement of the filopodia membrane. (c-e) Coiling shown by a schematic depiction (c), fluorescent image, scale bar is $5$~$\mu$m (d), and $3$D tracing of a filopodium from a MCF7 cell (cyan, EGFP Lifeact-7) (Supplementary Fig.~7) (e). Arrows in (e) indicate the binormals at different positions along the curve. (f) Schematic depiction of a rotating free filopodium. (g) Top: Confocal image of a filopodium extended from a MCF7 cell (white, EGFP Lifeact-7) overlayed with time color-coded images of the bead rotating around the actin shaft. Scale bar is $5$~$\mu$m. A vitronectin coated bead, indirectly attached to the actin fibers via transmembrane integrins, performs a spiral motion around the filopodium towards the cell body as a result of the internal twisting of actin shaft and retrograde actin flow (Fig.~2, and Supplementary Fig.~4 and 5). Bottom: Gabor-filtered $Z$-projection of the movement of a filopodial tip of an EGFP Lifeact-7 labelled MCF7 cell at consecutive color-coded time points. Scale bar is $1$~$\mu$m (Fig.~3).
Actin shaft twisting and tip rotation occur at similar rates. (h) Tip tracking in the $YZ$~plane reveals counterclockwise rotation as seen from the tip towards the cell body. (i) Schematic depiction of an optically trapped filopodium performing traction due to buckling and thus apparent shortening induced by accumulation of twist. (j) Fluorescent image of an optically trapped and buckling filopodium extracted from a HEK293T cell (cyan, EGFP Lifeact-7) using a $4.95$~$\mu$m bead. Scale bar is $5$~$\mu$m.
(k) Force profile from a filopodium extracted from a HV.5.1 mESC in S/L medium.

\clearpage
\bibliographystyle{apsrev}
\bibliography{main.bib}

\end{document}